\documentclass{aa}  
\usepackage{graphicx}
\usepackage{txfonts}
\def\deg{$^\circ$\hspace{-1.3mm}.\hspace{0.3mm}}  
\def\as{\arcsec\hspace{-1.2mm}.\hspace{0.3mm}} 
\def\am{\arcmin\hspace{-1.2mm}.\hspace{0.3mm}}  
\def\e{e$^-$}

\begin{document}
   \title{UBVRI twilight sky brightness at ESO-Paranal \thanks{Based
   on observations made with ESO Telescopes at Paranal Observatory.}}

   \subtitle{}

   \author{F. Patat\inst{1}
   \and
    O.S. Ugolnikov\inst{2}
   \and
    O.V. Postylyakov\inst{3}}

   \offprints{F. Patat}

   \institute{European Southern Observatory (ESO), K. Schwarzschildstr. 2,
              D-85748, Garching b. M\"unchen, Germany\\
              \email{fpatat@eso.org}
              \and
              Space Research Institute, Russian Academy of Sciences, 
              Profsoyuznaya ul., 84/32, Moscow, 117997 Russia\\
              \email{ugol@tanatos.asc.rssi.ru}
              \and
              A.M.Obukhov's Institute of Atmospheric Physics, Russian
              Academy of Sciences, Pyzhevsky per., 3, Moscow 119017 Russia\\
              \email{ovp@ifaran.ru}
             }

   \date{Received ...; accepted ...}

 
  \abstract
   {Twilight studies have proved to be important tools to analyze the 
   atmospheric structure with interesting consequences on the characterization
   of astronomical sites. Active discussions on this topic have been recently 
   restarted in connection with the evaluation of Dome C, Antarctica as a potential 
   astronomical site and several site-testing experiments, including twilight brightness 
   measurements, are being prepared.}
   {The present work provides for the first time absolute photometric 
   measurements of twilight sky brightness for ESO-Paranal (Chile), which are 
   meant both as a contribution to the site monitoring and as reference 
   values in the analysis of other sites, including Dome C.}
   {The $UBVRI$ twilight sky brightness was estimated on more than 2000 FORS1
   archival images, which include both flats and standard stars observations 
   taken in twilight, covering a Sun zenith distance range 
   94$^\circ$-112$^\circ$.}
   {The comparison with a low altitude site shows that Paranal $V$ twilight
   sky brightness is about 30\% lower, implying that some fraction of multiple
   scattering has to take place at an altitude of a few km above the sea
   level.}
   {}


   \maketitle
%

\section{\label{sec:intro}Introduction}

The quality of an astronomical site is determined by several
parameters, which may vary according to the wavelength range of
interest. For the optical and near-IR domain, these include typical
seeing, sky transparency, number of clear nights, humidity, night sky
brightness, amount of precipitable water vapor, dust and
aerosols. While the seeing, extinction, sky brightness and other
quantities are commonly measured at most observatories, the twilight
brightness is not. This is mainly because the relevant information on
the typical atmospheric conditions can be derived from other
measurements obtained during the night. Nevertheless, twilight
observations provide an independent tool for probing the overhead
atmosphere under much higher flux conditions, thus allowing more
accurate results. The interested reader can find an extensive review
on this topic in the classical textbook by Rozenberg
(\cite{rozenberg}).

Very recently, the twilight has received particular attention due to
the growing interest of the international astronomical community for
what seems to be the new frontier of ground-based astronomy, i.e. Dome C -
Antarctica. This site is exceptional under many respects. Besides the
extremely good seeing conditions reported by Lawrence et
al. (\cite{lawrence}), several studies have shown very low amounts of
precipitable water vapor which, coupled to a low sky emission, could
imply that this is the best site for IR and sub-millimetric
ground-based astronomy. A review about the characteristics of Dome C
has been recently presented by Kenyon \& Storey (\cite{kenyon}).

One of the main concerns is related to the high latitude of this
site. This, in fact, causes a significant reduction in the amount of
dark time with respect to equatorial observatories, thus posing some
doubts about the effective exploitation of the exceptional seeing in
the optical. The possibility of opening spectral windows otherwise
unaccessible from the ground is in itself a valid and sufficient
scientific driver for Dome C. Nevertheless, arguments in favor of Dome
C as a site for optical astronomy have been put forward. Among these,
a smaller average contribution by the scattered moonlight to the
global background and a cleaner atmosphere have been advocated as
features that may possibly compensate for the reduced dark time
(Kenyon \& Storey
\cite{kenyon}). In particular, since the last phases of twilight (also
known as deep twilight) are dominated by multiple scattering
(Rozenberg \cite{rozenberg}), the amount of scattered sunlight is
strongly dependent on the amount of aerosol in the lower atmospheric
layers (see for example Ougolnikov et al. \cite{oug04}). In a
supposedly low aerosol content site like Dome C, this effect is
expected to be very low and this, in turn, would allow one to start
the observations earlier than at {\it normal} sites. Even though the
argument has a good physical ground, direct on site measurements are
still lacking. In this respect, it is worth mentioning that a couple
of dedicated experiments for sky brightness measurements are currently
being setup (A. Moore, J. Storey 2006, private communications).

In spite of the large number of investigations done in the past in
this field, absolute twilight brightness measurements are rather rare,
especially for large observatories placed in top rated sites. To our
knowledge, the only published work on twilight observations in the
Johnson-Cousins standard system is the one by Tyson \& Gal
(\cite{tyson}) who however, given their purposes, report only
uncalibrated data for CTIO. In the light of these fact, both with the
purpose of characterizing Paranal also from this new point of view and
to provide the community with absolute reference values obtained over
a large time baseline, we present here for the first time $UBVRI$
twilight sky brightness measurements.

The paper is organized as follows. In Sec.~\ref{sec:model} we
introduce the basic concepts through a simplified model (which is
discussed in more detail in Appendix A), while in Sec.~\ref{sec:obs}
we describe the observations, data reduction and calibration. The
$UBVRI$ twilight sky brightness measured at ESO-Paranal is presented
and discussed in Sec.~\ref{sec:results}, while Sec.~\ref{sec:discuss}
summarizes the results obtained in this work.

\section{\label{sec:model}The twilight problem}

The calculation of scattered flux during twilight is a rather
complicated problem that requires a detailed treatment of multiple
scattering (see for example Bl\"attner et al. \cite{blattner} and Wu
\& Lu \cite{wu}) and an accurate description of the atmospheric
composition and the physical phenomena taking place in the various
layers (Divari \& Plotnikova \cite{divari}; Rozenberg
\cite{rozenberg}). Notwithstanding the large amount of work done in
the '60 and in the '70, the problem is still matter of investigations
(see for example Anderson \& Lloyd
\cite{anderson}; Ougolnikov \cite{oug}; Ougolnikov \& Maslov
\cite{oug02}; Ekstrom \cite{ekstrom}; Ougolnikov, Postylyakov \& Maslov 
\cite{oug04}; Postylyakov \cite{postylyakov}; Mateshvili et al. 
\cite{mateshvili}). While it is well beyond the purposes of the present work 
to explore the problem from a theoretical point of view, we deem it is
interesting to introduce a simple single-scattering model, which is
useful both to understand the basic principles of twilight and to
provide a quick comparison to the observed data. The assumptions and
the model itself are discussed in Appendix \ref{sec:appendix}, to
which we refer the interested reader for the details, while here we
concentrate on the model predictions only.

\begin{figure}
\centering
\includegraphics[width=8cm]{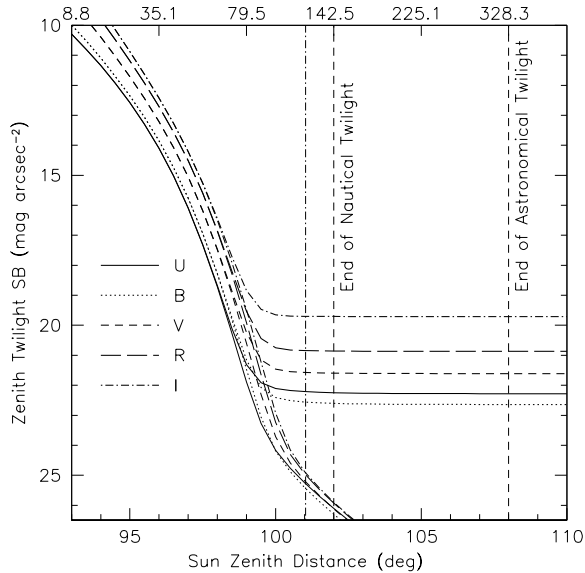}
\caption{\label{fig:model}Model twilight sky brightness at zenith. The thick 
curves include the night sky contribution, while the thin lines
indicate the scattered component only.  The vertical dashed-dotted
line marks the Sun zenith distance when the lower boundary layer height is
120 km.  The upper scale indicates the lower Earth's boundary layer
height in km.}
\end{figure}

The calculated zenith ($\alpha$=0) $UBVRI$ sky brightness as a
function of Sun zenith distance $\zeta$, computed using the average
broad band extinction coefficients for Paranal (Patat
\cite{patat03a}), are plotted in Fig.~\ref{fig:model}. 
As one can immediately see, the single scattering component drops
below the night sky brightness between $\zeta$=99$^\circ$ and
$\zeta=$100$^\circ$, indicating that from this point on multiple
scattering is the only contributor to the observed flux, as shown by
Ugolnikov \& Maslov (\cite{oug02}) on the basis of polarization
measurements.

Another aspect to be considered is the fact that the transition to the
flatter part of the atmospheric density profile (see
Fig.~\ref{fig:msis90}) definitely occurs during the multiple
scattering-dominated phase. Since multiple scattering takes place with
higher probability where the density is higher, i.e. in the lower
atmospheric layers, the explanation given by Tyson \& Gal
(\cite{tyson}) for the observed brightness decline rate during
twilight does not seem to be correct.  In fact, these authors
interpret the observed values as the pure consequence of the lower
shadow boundary height change, neglecting extinction and multiple
scattering. They conclude that, since their observations have been
taken when 100$\leq h_z \leq$ 400 km (where $h_z$ is the height of the
lower Earth's shadow boundary along the zenith direction; see also
Appendix A), the sky brightness rate is directly related to the slope
of the density law in that region of the atmosphere.  Nevertheless,
the calculation of Sun's ephemeris for the site and epoch of Tyson \&
Gal's observations shows that, in the most extreme case (see their
Table~1, $R$ filter), it was 1\deg8 $\leq \varphi \leq$ 7\deg4
(where $\varphi=\zeta-90$).  As the reader can figure out from
Table~\ref{tab:height}, this implies that $h_z<$80 km in all cases,
i.e.  well within the steep part of the density profile. Therefore,
the fact that the observed rate and the rate expected from pure single
scattering in the higher atmospheric layers are consistent, is just a
coincidence.

\section{\label{sec:obs}Observations, Data Reduction and Calibration}

In order to measure the twilight sky brightness on Paranal, we have
used archival calibration data obtained with the FOcal Reducer/low
dispersion Spectrograph (hereafter FORS1), mounted at the Cassegrain
focus of ESO--Antu/Melipal 8.2m telescopes (Szeifert 2002). The
instrument is equipped with a 2048$\times$2048 pixels (px) TK2048EB4-1
backside thinned CCD and has two remotely exchangeable collimators,
which give a projected scale of 0\as2 and 0\as1 per pixel (24$\mu$m
$\times$ 24$\mu$m). According to the used collimator, the sky area
covered by the detector is 6\am8$\times$6\am8 and 3\am4$\times$3\am4,
respectively.  For this study we have selected only the data obtained
with the lower resolution collimator and the 4-port high-gain read-out
mode, since this combination is the most used for imaging with
FORS1. With this setup the read-out noise is 5.5 electrons (\e).

\begin{figure}
\centering
\includegraphics[width=8cm]{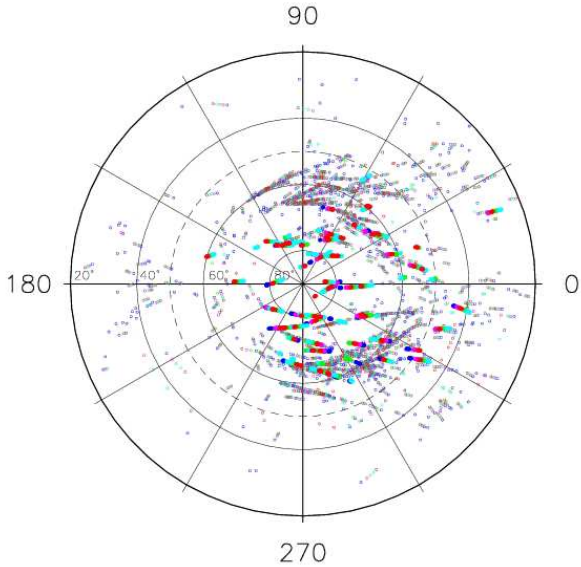}
\caption{\label{fig:altaz}Distribution of twilight observations in Alt-Az
coordinates for TSF (filled symbols) and LSE (empty symbols). The
astronomical azimuth has been replaced with the difference in azimuth
between the telescope pointing and the Sun, $\Delta a_\odot$.}
\end{figure}

For our purposes, we have selected two sets of data. The first is
composed by broad-band $UBVRI$ twilight sky flats (hereafter TSF),
which are regularly obtained as part of the calibration plan. In the
current implementation, after taking a test exposure the observing
software estimates with a simple algorithm the integration time for
the first exposure in a series of 4 frames. Subsequent exposure times
are adjusted on the basis of the previous exposure level and this
allows one to obtain high signal-to-noise images with a rather
constant counts level, which is typically around 20,000 ADUs. This is
achieved with exposure times which range from 0.25 sec up to 5
minutes. Given these values, the sensitivity of FORS1 in the various
passbands and the typical twilight sky brightness behaviour (see for
example Fig.~\ref{fig:model}), these observations are expected to
approximately cover the Sun zenith distance range 94$^\circ$ $\leq
\zeta
\leq$ 101$^\circ$, i.e. still within the nautical twilight. Since the
most important part of this analysis concerns the deep twilight, it
is clear that an additional set of data is required to complement the
sky flats. 

The calibration plan of FORS1 includes the observation of standard
stars fields (Landolt \cite{landolt}) in $UBVRI$ passbands, which are
regularly taken during twilight, typically just after the sky flats
sequence is completed. For calibration purposes, a fraction of these
exposures are obtained using relatively long integration times
(typically 40 sec for $U$ and 20 sec for $BVRI$) which, at an 8
m-class telescope, are sufficient to bring the sky background at
exposure levels which are suitable to our purposes. In fact, the bulk
of these observations covers the range 100$^\circ$ $\leq \zeta
\leq$ 112$^\circ$, i.e. well into astronomical twilight. 
For the sake of clarity we will indicate them as long exposure
standards (LES).

\begin{figure}
\centering
\includegraphics[width=8cm]{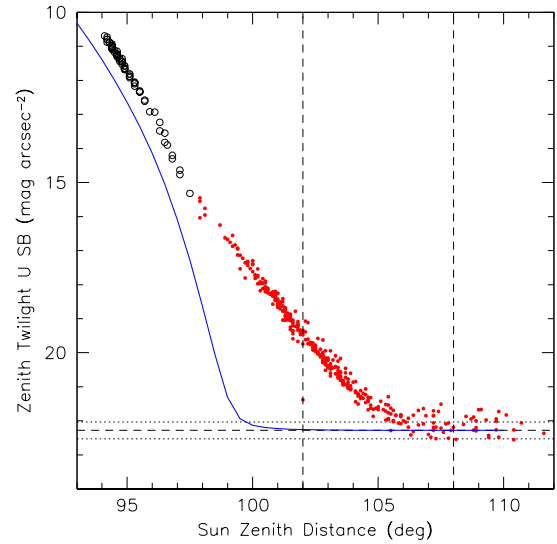}
\caption{\label{fig:twiU}Zenith twilight sky surface brightness in the 
$U$ passband from TSF (empty symbols) and LSE (filled symbols) with
$|\alpha|\leq$40$^\circ$.  The vertical dashed lines mark the end of
nautical (left) and astronomical (right) twilight. The solid curve
traces the simple model described in the text.  The horizontal lines
indicate the $U$ dark time night sky brightness average value (dashed)
for Paranal (Patat \cite{patat03a}) and the $\pm$1$\sigma$ levels (dotted).}
\end{figure}

In order to collect a statistically significant sample, we have
retrieved from the ESO Archive all suitable TSF obtained in $UBVRI$
passbands from 01-01-2005 to 30-09-2005, for a total of 1083 frames
($U$: 148, $B$: 208, $V$: 226, $R$: 261, $I$: 240). Since a much higher
night-to-night spread is expected in the deep twilight phase due to
the natural fluctuations of the night sky emission and also because
LSE are less frequently obtained than TSF, a larger time interval
must be considered. To this aim, we have put together the LES sample
collecting all suitable images obtained from 01-01-1999 (i.e. shortly
after the beginning of FORS1 operations) to 30-09-2005, thus covering
almost 6 years, for a total of 3388 frames ($U$: 923, $B$: 611, $V$: 609, $R$:
635, $I$: 610).  All images were processed within the {\tt xccdred}
package of IRAF\footnote{IRAF is distributed by the National Optical
Astronomy Observatories, which are operated by the Association of
Universities for Research in Astronomy, under contract with the
National Science Foundation.}. Due to the large amount of data and the
purpose of this work, the bias subtraction was performed using a
pre-scan correction only, while flat-fielding was achieved using a
stack of all TSF in each given passband as a master flat, which was
then adopted to correct each TSF and LES frame.

Due to the nature of the data, there is no need for taking care of the
possible presence of crowded stellar fields or bright extended
objects, as it is the case for night sky brightness estimates (see
Patat \cite{patat03b}).  For this reason, the background in each image
was measured using a simple and robust mode estimator. To avoid
possible vignetting and flat fielding problems, only the central
1024$\times$1024 pixels were considered. On these spatial scales and
due to improper flat-fielding, FORS1 is known to show variations of
the order of a few percent, while the gradients in the twilight sky
are much smaller (see Chromey \& Hasselbacher
\cite{chromey}) and can be safely neglected. Therefore, the mode $<I>$ of 
pixel intensity distribution is assumed as the best estimate of the
sky background. For each filter, this is converted into a surface
brightness in the Johnson-Cousins system via the following relation

\begin{displaymath}
b=-2.5 \log <I> + 2.5 \log (p^2 t_{exp}) + m_0
\end{displaymath}

where $p$ is the pixel scale (arcsec px$^{-1}$), $t_{exp}$ is the
exposure time (seconds) and $m_0$ is the instrumental photometric
zeropoint for the given passband. No color correction has been
applied, for two reasons: a) colour terms in FORS1 are very small
(Patat \cite{patat03a}); b) color correction depends on the intrinsic
color which, for the twilight sky, changes according to time and
position.  Given this and the fact that night-to-night fluctuations
are the dominating source of noise in the measurements, the color
correction can be safely neglected. As for the instrumental zeropoints,
we have used the average values computed for the period of interest
using data taken during photometric nights only.  Variations in the
zeropoints are known to take place (Patat
\cite{patat03a}), mainly due to the aging of telescope reflective
surfaces but, again, these are smaller than the inherent sky
variations. Finally, no airmass correction has been applied and this
is expected to produce an additional spread in the data close to the
end of astronomical twilight and a systematic increase in the average
sky brightness in that region.  Due to the large number of pixels used
($N>$10$^6$), the typical internal photometric error is expected to be
less than 1\%.

\section{\label{sec:results}Twilight sky brightness at ESO-Paranal}

Not being obtained specifically for twilight brightness measurements,
the data are inhomogeneously distributed on the sky. In fact, in an
Alt-Az plot where the ordinary azimuth is replaced by the difference
in azimuth between the sky patch and the Sun ($\Delta a_\odot\equiv
a-a_\odot$), the data points tend to cluster in two regions, which
correspond to evening and morning observations (Fig.~\ref{fig:altaz}).
Besides target azimuth and altitude, for each data point we have
computed a series of other quantities, which are relevant for the
subsequent analysis. These include Sun azimuth and altitude,
Sun-target angular separation, Moon phase, Moon altitude and
Moon-target angular separation. To avoid contamination from scattered
moonlight in the LSE, we have selected only those data points for which
the Moon was below the horizon.

\begin{figure}
\centering
\includegraphics[width=8cm]{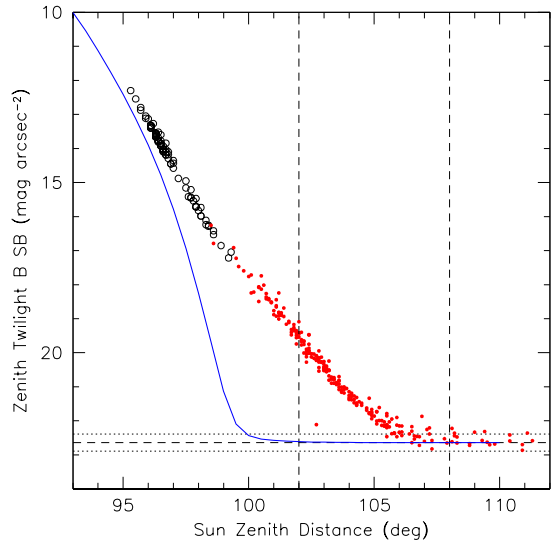}
\caption{\label{fig:twiB}Same as Fig.~\ref{fig:twiU} for the $B$ passband.}
\end{figure}

\begin{figure}
\centering
\includegraphics[width=8cm]{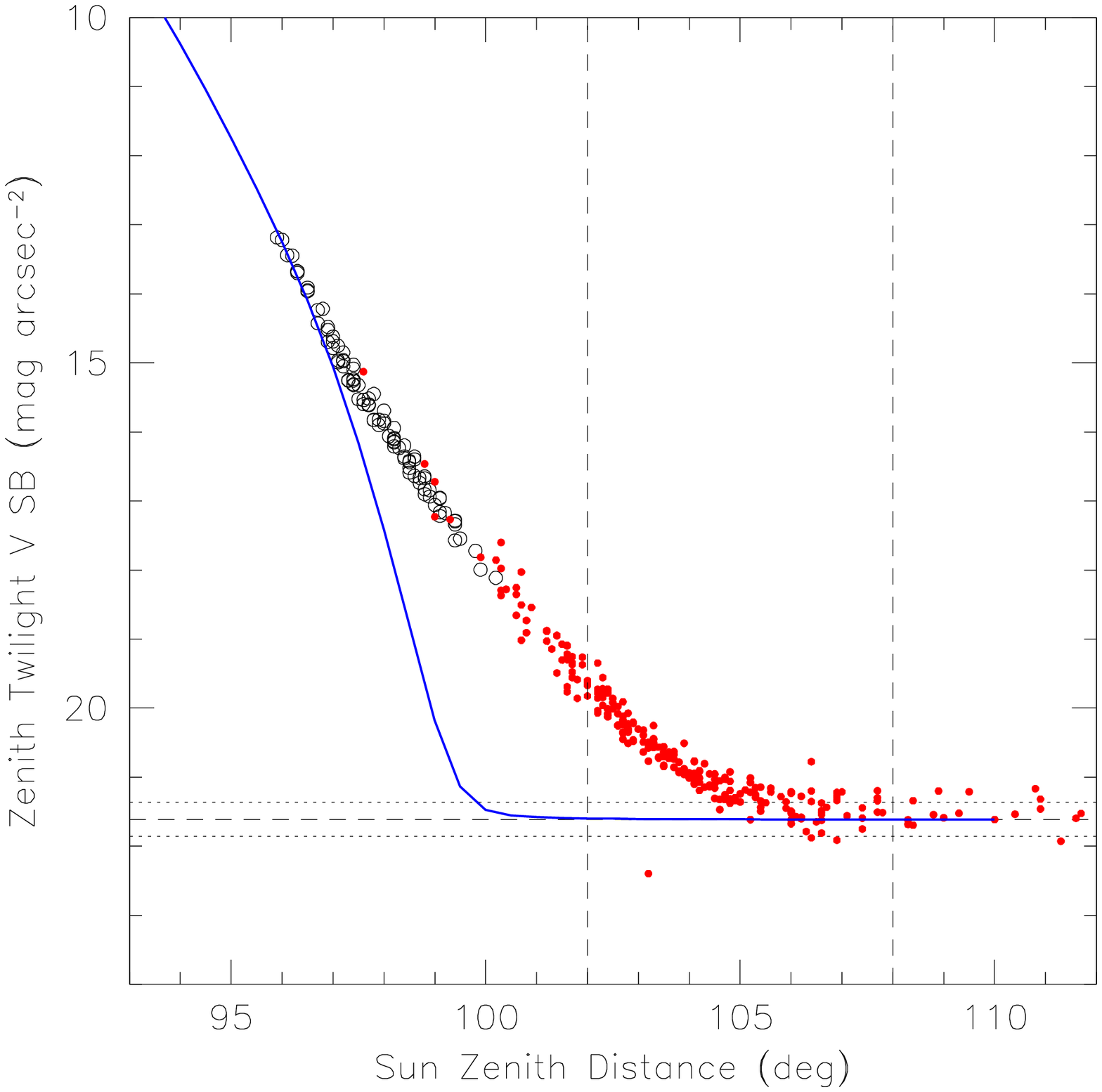}
\caption{\label{fig:twiV}Same as Fig.~\ref{fig:twiB} for the $V$ passband.}
\end{figure}

\begin{figure}
\centering
\includegraphics[width=8cm]{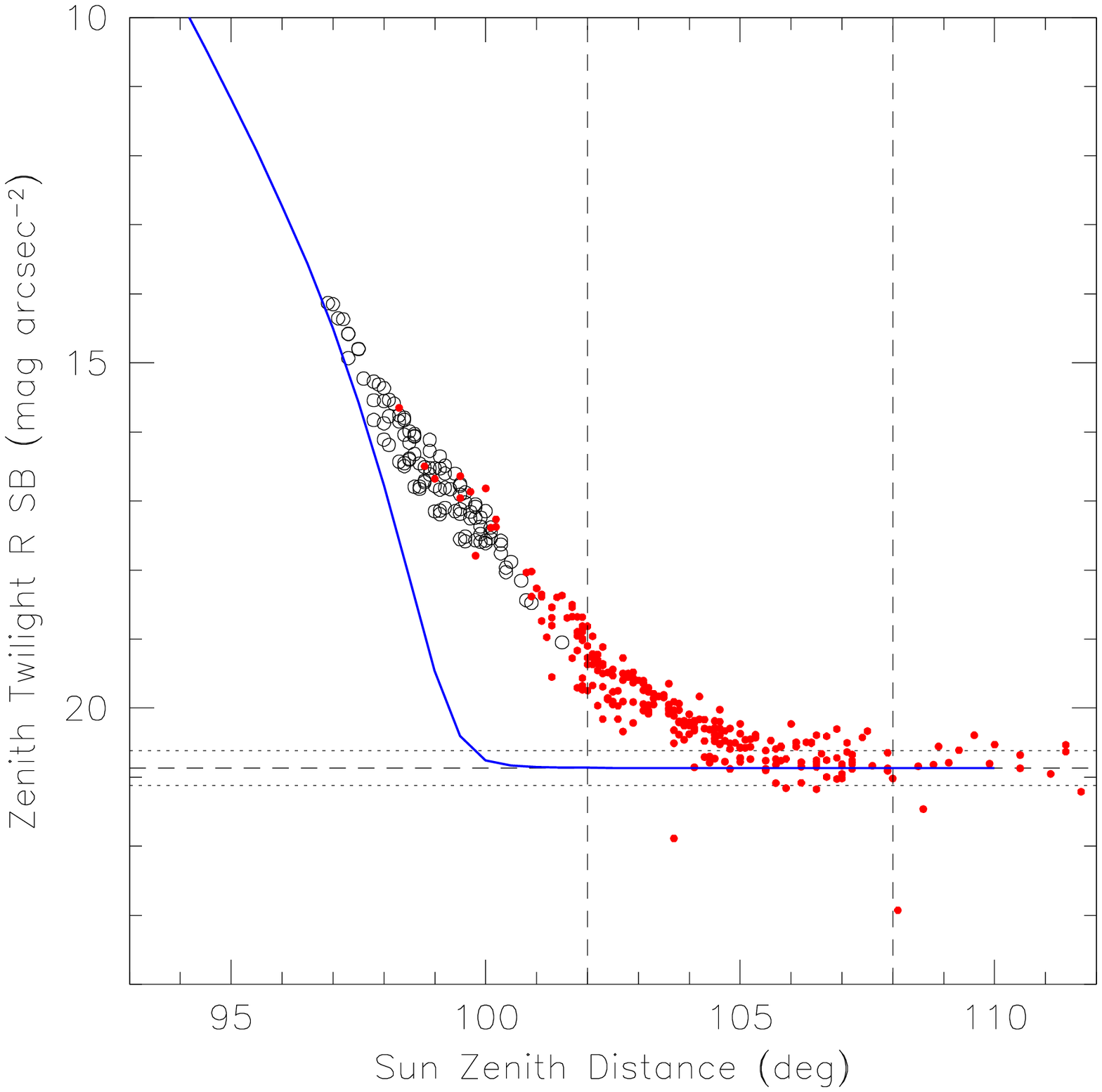}
\caption{\label{fig:twiR}Same as Fig.~\ref{fig:twiU} for the $R$ pass band.}
\end{figure}

\begin{figure}
\centering
\includegraphics[width=8cm]{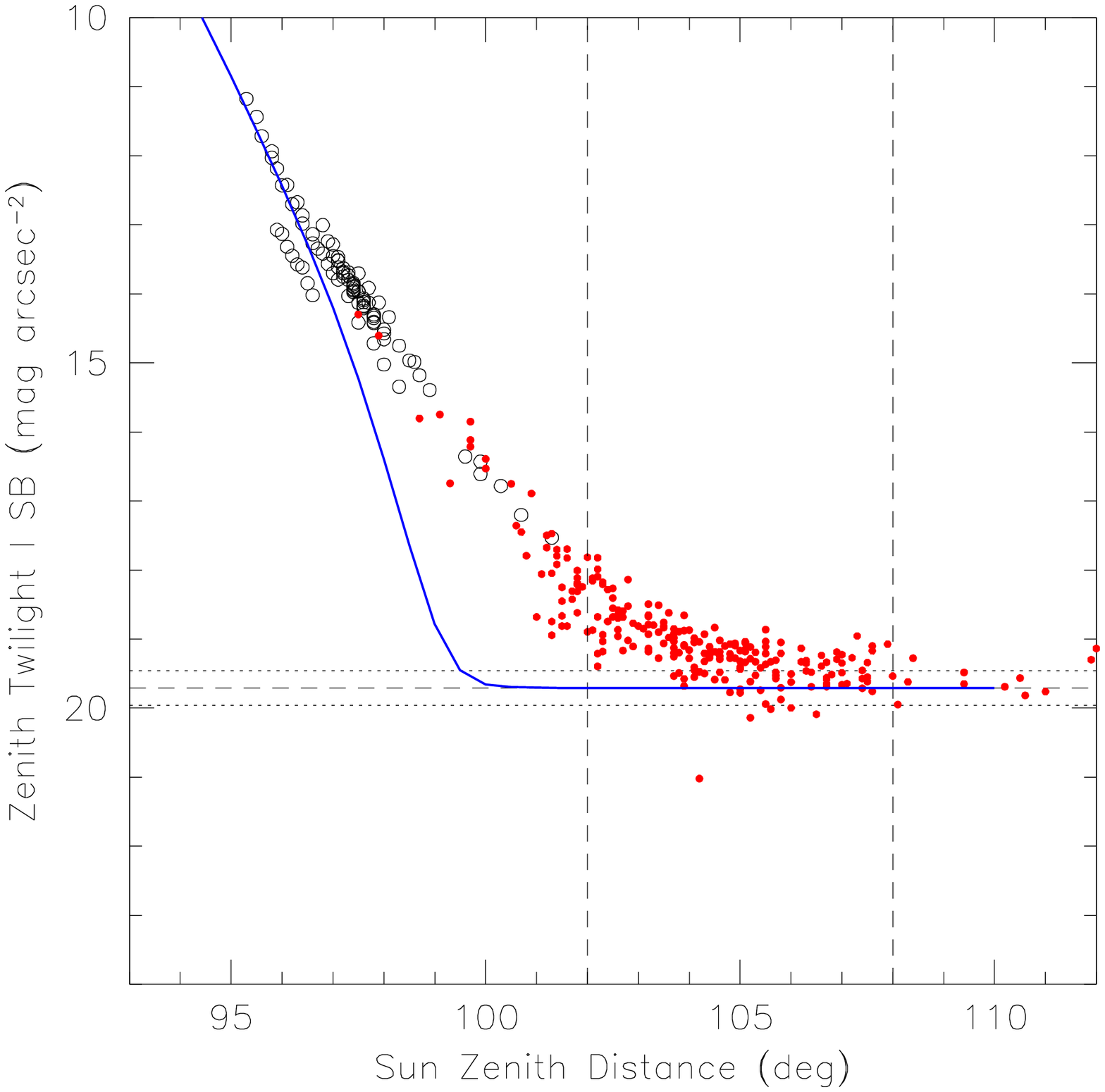}
\caption{\label{fig:twiI}Same as Fig.~\ref{fig:twiU} for the $I$ passband.}
\end{figure}

Since the twilight sky brightness for a given Sun zenith distance
changes with the position on the sky, in order to study its behavior
as a function of $\zeta$ it is necessary to make a selection on the
Alt-Az coordinates. Given the nature of the available data, which
appear to be rather concentrated (Fig.~\ref{fig:altaz}), it seems
reasonable to restrict the analysis to zenith region only. In order to
have a sufficient amount of measurements, we have used all data points
with zenith distance $|\alpha|\leq$40$^\circ$, which is of course
expected to cause some additional spread in the observed relation.
The results are presented in Figs.~\ref{fig:twiU}-\ref{fig:twiI} for
the $UBVRI$ passbands respectively.

\begin{table}
\caption{\label{tab:fit}Twilight sky brightness fitted parameters in the
range 95$^\circ \leq\zeta\leq$ 105$^\circ$. All values are expressed
in mag arcsec$^{-2}$.}
\center{
\begin{tabular}{cccccc}
\hline
Filter & $a_0$ & $a_1$ & $a_2$ & $\sigma$ & $\gamma$ \\
       &   & deg$^{-1}$ & deg$^{-2}$&  &  deg$^{-1}$\\
\hline
U      & 11.78 & 1.376 & $-$0.039 & 0.24 & 1.23$\pm$0.01\\
B      & 11.84 & 1.411 & $-$0.041 & 0.12 & 1.24$\pm$0.01\\
V      & 11.84 & 1.518 & $-$0.057 & 0.18 & 1.14$\pm$0.02\\
R      & 11.40 & 1.567 & $-$0.064 & 0.29 & 1.09$\pm$0.03\\
I      & 10.93 & 1.470 & $-$0.062 & 0.40 & 0.94$\pm$0.03\\
\hline
\end{tabular}
}
\end{table}

\begin{figure}
\centering
\includegraphics[width=8cm]{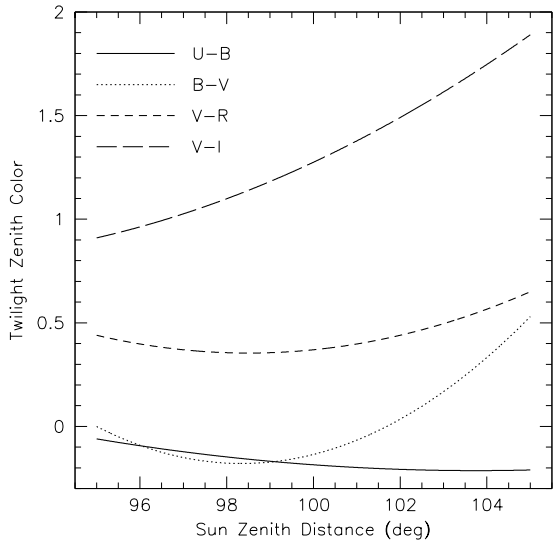}
\caption{\label{fig:color}Broad band zenith twilight sky colors. The curves
have been computed using second oder polynomials fitted to the
observed data.  For comparison, the colors of the Sun are $U-B$=0.13,
$B-V$=0.65, $V-R$=0.52 and $V-I$=0.81, while those of the night sky at
Paranal are $U-B$=$-$0.36, $B-V$=1.03, $V-R$=0.74 and $V-I$=1.90
(Patat \cite{patat03a}).}
\end{figure}

As expected, the single scatter model drops much faster than the
actual observations. Whilst for $VRI$ the model deviates from the data
around $\zeta\sim$97$^\circ$, for $B$ and especially for $U$ the model
underestimates the surface brightness already at
$\zeta\leq$96$^\circ$, indicating that multiple scattering is more
efficient at shorter wavelengths. This is in agreement with the
findings of Ougolnikov \& Maslov (\cite{oug02}), who have shown that
the contribution of single scattering in the phases immediately
following sunset is about 40\%, 60\%, 70\% and 80\% in $U$, $B$, $V$
and $R$ respectively. These fractions remain roughly constant until
$\zeta\leq$95$^\circ$, after which the role of single scattering
becomes weaker and weaker and multiple scattering takes rapidly
over. In all passbands, the night sky brightness level is reached at
around $\zeta$=105$^\circ$-106$^\circ$.

In order to give a more quantitative description of the observations,
we have fitted the surface brightness data in the range 95$^\circ \leq
\zeta \leq$ 105$^\circ$ using second order polynomials of the form
$a_0 + a_1 (\zeta-95) + a_2 (\zeta-95)^2$, with $\zeta$ expressed in
degrees and the surface brightness in mag arcsec$^{-2}$.  The results
are presented in Table~\ref{tab:fit}, where we have reported also the
RMS deviation from the fitted function $\sigma$ and the slope $\gamma$
deduced from a linear fit to the data in the range 95$^\circ
\leq \zeta \leq$ 100$^\circ$, i.e. during the interval typically used to obtain
TSF exposures, when the contribution by the night sky is still
moderate.  A first aspect to be noticed is the spread shown by the
data points around the mean laws, which are due to the night-to-night
variations in the atmospheric conditions. The dispersion becomes
particularly large in the $I$ passband, where the fluctuations appear
to be quite pronounced. As for the decay rate during nautical
twilight, we notice that this tends to decrease for increasing values
of wavelength.

To convert the values of $\gamma$ reported in
Table~\ref{tab:fit} into surface brightness variation per unit time,
one has to multiply them by $d \varphi/dt$, which is given by:

\begin{displaymath}
\frac{d\varphi}{dt} = \cos \varphi \sin H_\odot \cos \delta_\odot \cos \phi 
\;\frac{dH_\odot}{dt}
\end{displaymath}

where $H_\odot$ and $\delta_\odot$ are the hour angle and declination
of the Sun respectively, while $\phi$ is the site latitude (see for
example Smart
\cite{smart}). Since $d H_\odot/dt\simeq$0.25 deg min$^{-1}$, for $\varphi 
\sim$0 (i.e. at the time of sunrise and sunset) for Paranal ($\phi$=$-$24\deg6)
one obtains $d \varphi/dt$=0.23 deg min$^{-1}$ at the equinoxes
($\delta_\odot$=0). Applying this factor to the values reported in the
last column of Table~\ref{tab:fit}, one obtains brightness decline
rates that range from 0.28 ($U$) to 0.22 ($I$) mag arcsec$^{-2}$
min$^{-1}$. These values can be directly compared with those reported
by Tyson \& Gal (\cite{tyson}). The data available to these authors
did not allow them to quantify the differences between the various filters
and hence they report a mean value of $\gamma$=0.23$\pm$0.02 mag arcsec$^{-2}$
min$^{-1}$ which is within the range defined by our data.

\begin{figure}
\centering
\includegraphics[width=8cm]{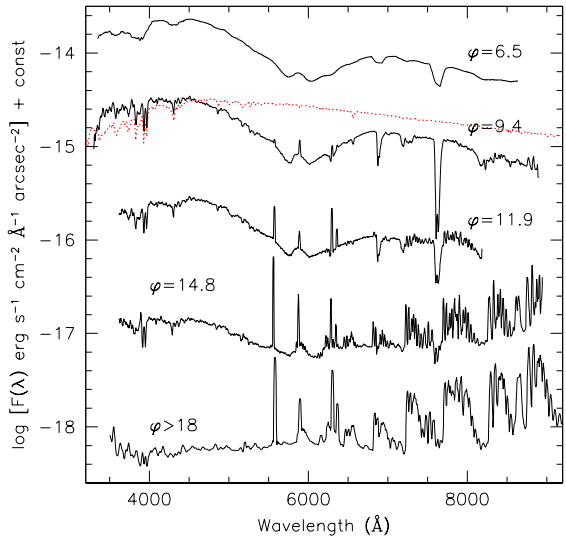}
\caption{\label{fig:spectra}Twilight spectra obtained at Paranal at different
Sun depression angles. For convenience, the spectra have been
corrected to the corresponding $V$ flux shown by the photometric
data. Spectral resolution and sky patch differ from spectra to spectra
(see Table~\ref{tab:spectra}).  For presentation the spectra have been
vertically shifted by the following amounts: +0.50 ($\varphi$=9.4),
+0.35 ($\varphi$=11.9), $-$0.10 ($\varphi$=14.8), $-$1.0 ($\varphi>$=18).
The dotted line traces the solar spectrum, normalized to the continuum
of the $\varphi$=9.4 spectrum at 4500 \AA.}
\end{figure}

With the aid of the second order best fit relations, we have computed
the color curves presented in Fig.~\ref{fig:color}. Due to the
dispersion of the observed data, these colors have to be regarded only
as indicative, especially in the region $\zeta>$102$^\circ$, where the
inherent night sky brightness fluctuations start to be significant,
particularly in the red passbands. It is interesting to note that
while $U-B$ and $V-R$ colors do not change very much as the Sun sinks
below the horizon, significant changes take place in $B-V$ and
especially in $V-I$.  In principle one expects that since multiple
scattering boosts the light at shorter wavelengths with respect to the
pure single scattering component, the overall color gets bluer and
bluer as the Sun deepens below the horizon. Then, at some point, the
night sky glow which has completely different colours, starts to
contribute and the colors progressively turn to those typical of the
night sky. As a matter of fact, the observed $U-B$, $B-V$ and $V-R$
curves indeed show this behaviour (see Fig.~\ref{fig:color}), while
$V-I$ turns steadily redwards. This is due too the interplay between
input Sun spectrum, scattering efficiency, extinction, multiple
scattering and the emergence of the night glow, which combine together
in quite a complicated way.  This is clearly illustrated in
Fig.~\ref{fig:spectra}, where we present twilight spectra obtained on
Paranal with FORS1 and his twin instrument FORS2. The sky spectra were
extracted from spectrophotometric standard stars observations taken
during twilight (see Table~\ref{tab:spectra}) and were wavelength and
flux calibrated with standard procedures in IRAF. The exposure times
ranged from 10 sec to 120 seconds and the signal-to-noise ratio was
increased meshing all pixels in the direction perpendicular to the
dispersion, after removing the region of the detector affected by the
well exposed standard star spectrum. For comparison,
Fig.~\ref{fig:spectra} shows also the typical dark time spectrum of
Paranal (Patat \cite{patat03a}) and the solar spectrum.

With the exception of the first spectrum, which was obtained with a
very low resolution ($\sim$130 \AA\/ FWHM), the remaining data allow one
to detect quite a number of details. For $\varphi<$12$^\circ$,
i.e. during the nautical twilight, the spectrum is rather different
from that of the Sun, even though it shows clear solar features, like
the CaII H\&K lines and the G-band at about 4300 \AA. The Rayleigh
scattered Sun flux gives a clear contribution in the region
bluewards of 5000\AA\/ down to $\varphi$=15$^\circ$, after which the
pseudo-continuum of the night sky emission takes over.

For $\varphi<$9$^\circ$, the contribution by night sky emission lines
is very weak. Characteristic lines like [OI]5577\AA\/ and [OI]6300,6364
\AA\/ (the latter overimposed on a O$_2$ absorption band) are barely 
visible, while the OH Meinel bands start to appear above 8000 \AA. One
remarkable exception are the NaI D lines, which are known to be
present during the so called {\em sodium flash} (Rozenberg
\cite{rozenberg}). A similar phenomenon is present for the
[OI]6300,6364 \AA\/ doublet (see for example Roach \& Gordon
\cite{gordon}), which is indeed visible in Fig.~\ref{fig:spectra}
already at $\varphi$=9\deg4.  With the onset of astronomical twilight,
the spectrum in the red is more and more dominated by the OH bands.
Another remarkable fact is the behaviour of the H$_2$O (6800, 7200
\AA) and O$_2$ (7600 \AA) molecular absorption bands. During the
bright twilight, when single scattering is still relevant
($\varphi$=6\deg5), they already appear to be significant, but they
become even deeper in the multiple scattering-dominated phase
($\varphi$=9\deg4), due to the longer optical path traveled by the
multiply scattered photons. For higher values of $\varphi$ they
progressively disappear due to the weakening of the scattered Sun's
continuum.

\begin{table}
\caption{\label{tab:spectra}Basic data for twilight sky spectra shown in
Fig.~\ref{fig:spectra}.}
\center{
\begin{tabular}{cccccc}
\hline
Date       & UT start & $\Delta \lambda/\lambda$ & Alt & Az & $\varphi$ \\
           &              & \AA (FWHM)               & deg & deg & deg \\
\hline
2001-09-21 & 10:06:58 & 130 & 82.6 &117.5 & 6.5 \\ 
2003-11-29 & 09:02:25 & 13 & 65.8 &122.9 & 9.4 \\ 
2003-02-28 & 09:45:52 & 16 & 27.5 &299.3 & 11.9 \\ 
1999-04-16 & 09:54:36 & 12 & 41.1 &233.4 & 14.8 \\
\hline
\end{tabular}
}
\end{table}

\begin{figure}
\centering
\includegraphics[width=8cm]{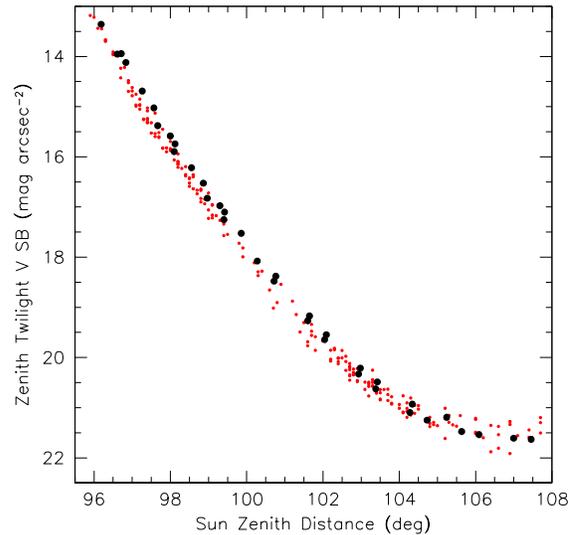}
\caption{\label{fig:comp} Comparison between the zenith twilight sky brightness
measured on Paranal (small symbols) and Crimean Astrophysical
Observatory (large symbols) for the $V$ passband.}
\end{figure}

In order to estimate the effects of site altitude on the twilight sky
brightness, we have compared the results presented here with data
obtained at the Southern Laboratory of the Sternberg Astronomical
Institute (Moscow, Russia) during three morning twilights on December
9-11, 2002 (see Fig.~\ref{fig:comp}). This facility is located
within the Crimean Astrophysical Observatory (CrAO), at a latitude of
44\deg7 North and 600 m a.s.l.. The observations were performed using
a wide-field CCD-camera with a field of view of
8$^\circ\times$6$^\circ$ and exposure times that ranged from 0.01 to 18
seconds. Photometric calibration was achieved using field stars
included in the Tycho-2 Catalog (Hog et al. \cite{hog}).  The
photometric passband of this instrument is fairly similar to the
Johnson-Cousins $V$, with a colour correction of the order of 0.01 mag
for the $(B-V)$ colour range shown by twilight data.

The two data sets clearly show that the twilight background at CrAO is
systematically brighter; the difference is constant during the dark
twilight period and vanishes at nightfall. More precisely, the
comparison between V band data in moonless conditions at CrAO (zenith)
and ESO-Paranal ($|\alpha|\leq 20^\circ$) shows that the mean
difference in the Sun depression range 5\deg5 $\leq \varphi
\leq$ 11\deg0 is $\Delta V$ = 0.27$\pm$0.03. On the other hand,
the typical atmospheric pressure value for ESO-Paranal is $P_1$=743
hPa, and for CrAO during the observations $P_2$=961
hPa. Interestingly, the magnitude difference implied by the pressure
ratio at the two sites is $-2.5\log (P_1/P_2)$=0.28, which is indeed
very similar to the measured difference $\Delta V$.  Therefore, we can
conclude that the deep twilight sky brightness is proportional to the
atmospheric pressure or, equivalently, to the atmospheric column
density above the observer. In turn, this implies that the light
undergoes multiple scattering throughout the whole atmosphere and not
only in the upper layers. Given that the difference in altitude
between Paranal and CrAO is only 2 km, the observations we present
here suggest that some fraction of multiple scattering has to take
place in the first few km above the sea level.

\section{\label{sec:discuss}Discussion and Conclusions}
  
In this paper we have presented for the first time absolute $UBVRI$
twilight brightness measurements for the ESO-Paranal Observatory
(Chile) spanning almost 6 years. These measurements will serve as
reference values for the similar studies which will be soon conducted
at Dome C, Antarctica, as part of the site testing campaign. The
planned {\it in situ} spectrophotometric measurements will finally
clarify whether this exceptional location shows a lower aerosols
content, as expected both due to the icy soil and to its large
distance from the sea coast (Kenyon \& Storey \cite{kenyon}).

\begin{figure}
\centering
\includegraphics[width=8cm]{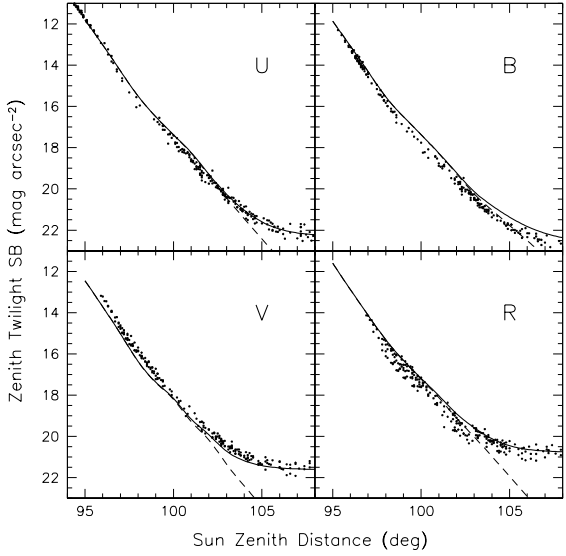}
\caption{\label{fig:compmcc}Comparison between Paranal zenith sky brightness
and the MCC++ model (Postylyakov \cite{postylyakov}). The dashed curve
traces the pure model solution without the contribution of night sky
emission.}
\end{figure}

The twilight sky brightness measurements presented here were obtained
from VLT-FORS1 archival data not specifically taken for this kind of
studies. Also, due to the large telescope diameter, the initial twilight
phases (0$^\circ \leq \varphi \leq$6$^\circ$) are not covered. In a
sense this is quite unfortunate, since for these low Sun depression
angles the lower shadow's boundary passes through the atmospheric
layers below $\sim$30 km (see Table~\ref{tab:height}), where the ozone
and aerosol stratospheric concentration is maximum. These phases are
in fact used to retrieve ozone and aerosol profiles, using both
intensity and polarization measurements (see for example Wu \& Lu
\cite{wu}; Ugolnikov et al. \cite{oug04}; Mateshvili et
al. \cite{mateshvili}). Nevertheless, during the deep twilight, when
the direct Sun radiation illuminates only the upper atmospheric layers
and single scattering on air molecules becomes progressively less
important, the amount of aerosols and ozone plays a relevant role
through multiple scattering.  Therefore, even though of much more
complicated interpretation, deep twilight observations may still give
some insights on the conditions in the lower atmosphere. Moreover, in
the context of the discussion about the supposedly shorter twilight
duration at Dome C (see Kenyon \& Storey \cite{kenyon}), what really
matters is the behavior during the deep twilight.  An example of this
kind of analysis is shown in Fig.~\ref{fig:compmcc}, where the data
obtained at Paranal are compared to the MCC++ model calculations
for a site at 2.6 km a.s.l. (see Postylyakov
\cite{postylyakov} for a detailed description). This code treats the 
radiative transfer in a spherical atmosphere including Ozone, Aerosol
and molecular scattering, also taking into account the backscattering
by the Earth's surface. As one can see the overall behavior is fairly
well reproduced. The deviations are possibly due to the differences in
real and model aerosol, since multiple scattering is very sensitive to
it. The model adopts a urban microphysical model for the first 10 km
of the atmosphere and this is certainly different from what is
expected for a desertic area close to the sea, as is the case of
Paranal.  Dedicated instruments for twilight sky brightness monitoring
coupled to detailed modeling may indeed give a useful contribution to
the already existing site testing tools, providing independent
indications about the overhead aerosol profile.

Some interesting results are obtained comparing the estimates obtained
for Paranal (2600 m a.s.l.) with those of a significantly lower site
like CrAO (600 m a.s.l.). Despite the fact that the bright twilight
and night sky brightnesses are very close at the two sites, during the
deep twilight Paranal is about 30\% darker than CrAO in the $V$
passband (see Fig.~\ref{fig:comp}). Due to the higher altitude,
Paranal suffers from a lower extinction  which, if all other
atmospheric properties are identical and multiple scattering takes
place mostly in the troposphere (5-10 km, Ougolnikov \& Maslov
\cite{oug02}), would then turn into a brighter twilight sky.
The observations actually show the opposite behaviour and the
brightness ratio is fairly consistent with the atmospheric pressure
ratio (see previous section). A natural interpretation of this fact is
that a fraction of the multiple scattering events takes place at heights which
are lower than it was originally thought, say below 3 km from the sea
level.

Whether this is due to the lower density of air molecules, to a
smaller amount of ground level aerosols or to a combination of the two
needs further investigation and the comparison with other astronomical
sites at even higher altitudes, like Mauna Kea.

\appendix

\section{\label{sec:appendix} A simple semi-analytical model for 
twilight brightness}

\subsection{Basic assumptions}

The model is based on the following simplifying assumptions: (i) Earth
is a sphere with radius $R_0$=6380 km; (ii) the atmosphere extends up
to $\Delta R$=400 km and the numerical density $n(h)$ of the
scattering particle density is given by the MSIS-E-90 model profile
(Hedin \cite{hedin}); (iii) the effect of atmospheric refraction can
be neglected; (iv) the Sun is a point source and all incoming sun rays
are parallel; (v) only single scattering plus attenuation is
considered; (vi) Rayleigh scattering by air molecules is the only
source of sunlight diffusion. While some of these assumptions are
reasonable, (v) and (vi) are a bit crude and will certainly lead to
discrepancies between the model result and the actual observations.
For a more sophisticated single scattering model taking into account
refraction and the presence of ozone and aerosols, the reader is
referred to Adams, Plass \& Kattawar (\cite{adams}).

\subsection{Model geometry}

\begin{figure}
\centering
\includegraphics[width=8cm]{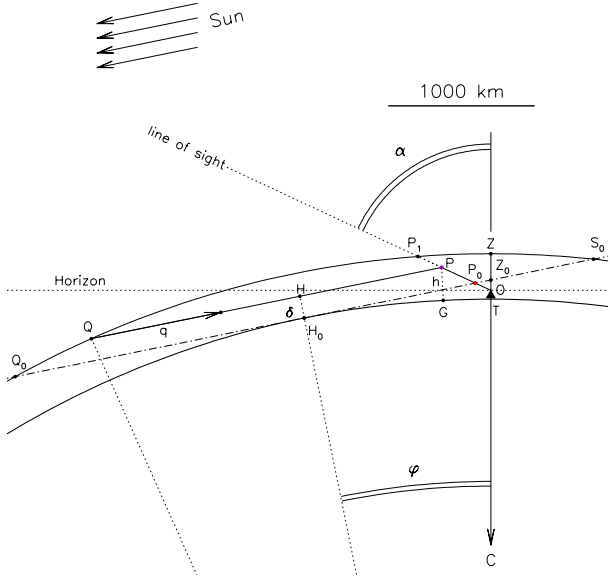}
\caption{\label{fig:geom}Geometry of the problem. For the sake of clarity, the
observer's elevation $h_s$ has been exaggerated.}
\end{figure}

As we have said, we will assume that the Earth is a sphere of radius
$R_0$ and that the observer is placed at an elevation $h_s$ above the
sea level. Since we will consider only small values of $h_s$ ($<$3
km), we make the further simplifying assumption that the horizon is a
plane tangent to the sphere of radius $R_0+h_s$ in $O$, (see
Fig.~\ref{fig:geom}), i.e. neglecting the horizon depression. We will
indicate with $\varphi$ the Sun depression ($\varphi>0$) and with
$\alpha$ the zenith distance of the generic sky patch under
consideration. For the sake of simplicity we will derive the sky
brightness only along the great circle passing through the zenith
($Z$) and the Sun. This angular distance is counted positively in the
direction of the Sun, so that negative angles indicate sky patches in
the anti-Sun direction. Under these simplifying assumptions, the lower
boundary of the Earth's shadow is described by a straight line, which
is tangent to the sphere in $H_0$, and which is indicated by a
dotted-dashed line in Fig.~\ref{fig:geom}. When the observer is
looking into the generic direction $\alpha$, the corresponding line of
sight will cross the lower shadow boundary in $P_0$ and the contribution to
the observed flux will come from all the scattering elements along the
segment $P_0 P_1$, where $P_1$ indicates the intersection between the
line of sight and the upper atmospheric boundary, which is placed at
an altitude $\Delta R$ above the sea level\footnote{The upper limit is
set just for numerical reasons.}.

In this geometry, a volume element placed in $P$ receives the sun
light, which is attenuated along its path $OP$, and it scatters the
photons into the observer's direction, with a scattering angle
$\theta=\pi/2-\alpha+\varphi$.  according to the scattering phase
function $\Phi(\theta)$, which obeys to the usual normalization
condition

\begin{equation}
\label{eq:norm}
\int_{4\pi} \Phi(\theta) \; d\Omega =1
\end{equation}

Before reaching the observer, it undergoes the extinction along
$OP$. The total flux will finally result from the integration along
the illuminated segment $P_0P_1$. In order to compute the required
quantities, we must first derive a series of useful relations. Since
one has to evaluate the particle density along the generic light path,
it is fundamental to know the height $h=GP$ above the sea level for any
given point along the trajectory. This is particularly simple for the
travel between $O$ and $P_1$, for which one can easily derive the
following relation:

\begin{equation}
\label{eq:hl}
h_{OP_1}=\sqrt{l^2  +2l(R_0+h_s)\cos\alpha + (R_0+h_s)^2} - R_0
\end{equation}

where $l$ is the coordinate along $OP_1$ ($l$=0 in $O$). Another
useful relation is the one that gives the lower limit for the integral
along the line of sight, i.e. the optical path $l_0$ between $O$ and
$P_0$. In order to get the required expression, one first needs to
find the length of $H_0P_0$ and $h_0$, i.e. the height above the sea
level of $P_0$. These can be obtained after some trigonometry and are
respectively

\begin{displaymath}
H_0P_0=R_0 \left [\tan \varphi - \frac{\sin \alpha}{\cos (\alpha-\varphi)} 
\left (\frac{1-\cos \varphi}{\cos \varphi} -\frac{h_s}{R_0}\right)
\right ] 
\end{displaymath}

and

\begin{displaymath}
h_0=\sqrt{H_0P_0^2 + R_0^2} - R_0
\end{displaymath}

Having these two equations at hand, one can easily find $l_0$:

\begin{displaymath}
\begin{array}{lll}
l_0 & = & \sqrt{(R_0+h_s)^2\cos^2 \alpha + h_0^2 -h_s^2+ 2R_0 (h_0-h_s)} \\
    &  & -(R_0+h_s)\cos \alpha
\end{array}
\end{displaymath}

As for the upper limit $l_1$ along the line of sight, this turns out to be:

\begin{displaymath}
\begin{array}{lll}
l_1 &=&\sqrt{(R_0+h_s)^2\cos^2 \alpha +2R_0(\Delta R-h_s) + \Delta R^2-h_s^2}\\
    & &-(R_0+h_s)\cos \alpha
\end{array}
\end{displaymath}

The next step is the calculation of $h_{QP}$, the height along the sun
ray. In order to do so, we introduce the perigee height $\delta$,
i.e. the minimum distance between the Earth's surface and the sun ray
passing through $P$, which can be expressed as a function of the $l$
coordinate along $P_0P_1$ as follows:

\begin{displaymath}
\delta=(l-l_0) \cos (\alpha-\varphi)
\end{displaymath}

Having this in mind, one can derive the length of $QH$ as follows:

\begin{displaymath}
QH=\sqrt{2R_0(\Delta R-\delta) + \Delta R^2 - \delta^2}
\end{displaymath}

If we then introduce a coordinate $q$ along $QP$ (with $q$=0 in $Q$),
considering the fact that $(QH-q)^2 + (R_0+\delta)^2=(R_0+h)^2$, we
finally obtain

\begin{equation}
\label{eq:hq}
h_{QP}=\sqrt{(QH-q)^2 + (R_0+\delta)^2} - R_0
\end{equation}

with $0\leq q \leq QP$. The optical path of the unscattered
sun ray $QP$ is given by the sum of $QH$ and $HP$, the latter being
$HP=H_0P_0-(l-l_0) \sin(\alpha-\varphi)$.  By means of Equations
\ref{eq:hl} and \ref{eq:hq} one can now compute the height above the
sea level (and hence the particle number density) along the light
path.The height along the zenith direction, $h_z=TZ_0$, can be readily
derived and it is given by

\begin{displaymath}
h_z=R_0 \frac{1-\cos \varphi}{\cos \varphi}
\end{displaymath}

From this one can figure out how fast $h_z$ grows when the sun
depression $\varphi$ increases (see also Tab.~\ref{tab:height}).
This, coupled to the rapid decrease of the atmospheric density as a
function of height, is the cause for the very quick drop in the
twilight sky surface brightness.

\begin{figure}
\centering
\includegraphics[width=8cm]{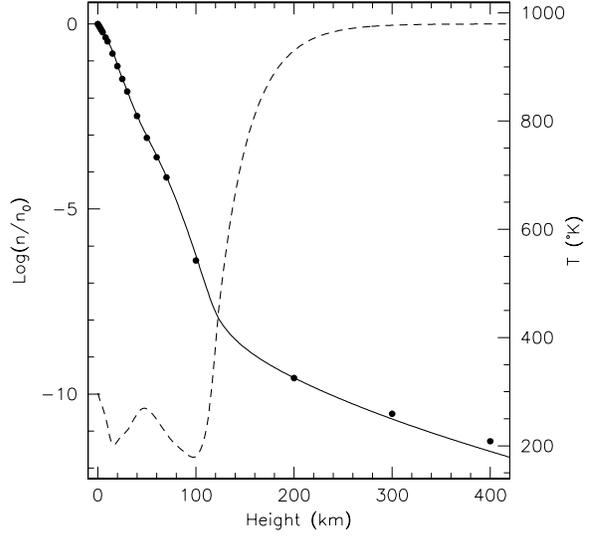}
\caption{\label{fig:msis90} Normalized density (solid line) and temperature
(dashed line) profiles according to the MSIS-E-90 model (Hedin
\cite{hedin}). For comparison, the dots trace the values of the US Standard
Atmosphere (McCartney \cite{mccartney}, Tab.~2.6).}
\end{figure}

\begin{table}
\caption{\label{tab:height} Height of lower boundary of Earth's shadow at 
zenith ($h_z$) and at 60$^\circ$ from zenith along the solar meridian
in the Sun's direction.}
\center{
\begin{tabular}{ccc}
\hline
$\varphi$ & $h_z$ & $h_0(\alpha=60)$ \\ (deg) & (km) & (km) \\
\hline
3       & 8.8   & 8.0  \\
6       & 35.1  & 29.9 \\
9       & 79.5  & 63.3 \\
12      & 142.5 & 106.7\\
15      & 225.1 & 159.1\\
18      & 328.3 & 220.1\\
\hline
\end{tabular}
}
\end{table}

\subsection{Density profile and optical depth}

As we have anticipated, for the density profile we have adopted the
MSIS-E-90 model\footnote{http://modelweb.gsfc.nasa.gov/models/msis.html} 
density profile (Hedin \cite{hedin}), which is presented in
Fig.~\ref{fig:msis90}.  As the plot shows, the global profile can be
roughly described by two laws: one exponential (for $h<$120 km, the so
called homosphere) and a power law (for $120<h<400$ km, the so called
thermosphere). Clearly, as the Sun depression increases, the lower
Earth's shadow boundary will pass through more and more tenuous
atmospheric layers and, therefore, the change in density slope should
turn into a change in the twilight sky surface brightness
decline. Under the assumptions made for this simplified model, this
should happen when $\varphi\simeq$101$^\circ$, i.e. close to the end
of nautical twilight.

Since in the following section we will be interested in the product
between the number density and the extinction cross section, we can
derive this, for a given passband, assuming the measured extinction
coefficient $\kappa(\lambda)$ (in mag airmass$^{-1}$) and integrating
the previous density profile along the vertical (i.e. at airmass
1). In fact, assuming that all the extinction comes from Rayleigh
scattering, one can write:

\begin{displaymath}
\tau_z(\lambda) = n_0 \; C_{ext}(\lambda) \int_{h_s} ^{\Delta R} 
\frac{n(h)}{n_0}  \; dh
\end{displaymath}

Then, considering that $\tau_z(\lambda)$=1.086$\kappa(\lambda)$, the
product $n_0 \; C_{ext}(\lambda)$ can be readily derived from the previous 
relation and used in all further optical depth calculations.

\subsection{Scattering cross section and phase function}

The $\lambda^{-4}$ wavelength dependency of the Rayleigh scattering
cross section is implicitly taken into account by the extinction
coefficients, which are to be considered as input data to the model
and not as free parameters. As for the scattering phase function we
have used the canonical expression for air molecules (McCartney
\cite{mccartney}):

\begin{displaymath}
\Phi(\theta) = \frac{3}{16\pi} [1+0.9324\cos^2(\theta)]
\end{displaymath}

where the multiplicative constant comes from the normalization
condition expressed by Equation \ref{eq:norm}.

\subsection{Scattered flux}

The scattered flux can be computed in the same way as done, for
example, by Krisciunas \& Schaefer (\cite{ks}) for the Moon light.  If
$L_\odot$ is the luminosity of the sun at a given wavelength
(expressed in photons per unit time and per unit wavelength), the flux
received by the Earth at the top of the atmosphere is $F_\odot^0 =
L_\odot/4\pi d^2$, where $d$=1 AU. If we now consider an infinitesimal
volume element $dV$ placed along the line of sight at a distance $l$
from $O$, the number of scattered photons received by the observer per
unit time, unit area and unit wavelength is given by

\begin{equation}
df = F_\odot^0 e^{-\tau(QP)}\; n[h(l)] \; C_{ext}(\lambda) 
\frac{\Phi(\theta)}{l^2}\;  e^{-\tau (OP)}\;dV
\end{equation}

Given the geometry of the problem, the infinitesimal volume element
can be written as $dV=dS \; dl\equiv \pi l^2 \phi^2 \; dl$, where $\phi$ is
the semi-amplitude of the angle subtended by $dS$ at the distance of
the observer. Since the solid angle subtended by $dS$ is $d\Omega=\pi
\phi^2$, the surface brightness produced by the volume element is simply

\begin{displaymath}
db=\frac{df}{d\Omega}= F_\odot^0  e^{-\tau(QP)} \; n[h(l)] \; 
C_{ext}(\lambda) \; \Phi(\theta) \; e^{-\tau(OP)} \; dl
\end{displaymath}

and the total surface brightness is finally obtained integrating along
the line of sight within the illuminated region:

\begin{displaymath}
b=\int_{l_0}^{l_1} db\; dl
\end{displaymath}

Finally, to take into account the contribution by the night sky
emission, we have added to the computed flux the one implied by the
average values measured for Paranal ($h_s$=2.6 km) and reported by
Patat (\cite{patat03a}, Tab.~4). With this, the model is completely
constrained and there are no free parameters.

\begin{acknowledgements}
We wish to thank S.L. Kenyon and J.W.V. Storey for inspiring this work
and the ESO Archive Group for the support received during the data
retrieval. O.Ugolnikov is supported by a Russian Science Support
Foundation grant. Our gratitude goes also to the referee,
Dr. A. Tokovinin, for his useful suggestions and comments.

This paper is based on observations made with ESO
Telescopes at Paranal Observatory.

\end{acknowledgements}


\begin{thebibliography}{}
\bibitem[1974]{adams} Adams, C.N., Plass, G.N. \& Kattawar, G.W., 1974,
        J. Atm. Sci., 31, 1662
\bibitem[1990]{anderson} Anderson, D.E. \& Lloyd, S.A., 1990, JGR, 95, 7429
\bibitem[1974]{blattner} Bl\"attner, W.G., Horak, H.G., Collins, D.G. \&
	Wells, M.B., 1974, Applied Optics, 13, 534
\bibitem[1996]{chromey} Chromey, F.R. \&
        Hasselbacher, D.A. 1996, \pasp, 108, 944
\bibitem[1966]{divari} Divari, N.B. \& Plotnikova, L.I., 1966, 
	Sov. Astron., 9, 840
\bibitem[2002]{ekstrom} Ekstrom, P., 2002, SPIE, 4815-14
\bibitem[1973]{gordon} Roach, F.E. \& Gordon J.L, 1973, The light of 
	the night sky, (Boston, Dordrecht Reidel)
\bibitem[1991]{hedin} Hedin, A.E. 1991, J. Geophys. Res. 96, 1159
\bibitem[2000]{hog} Hog, E. et al., 2000, The Tycho-2 Catalogue on CD-ROM,
        Copenhagen University Observatory
\bibitem[2005]{kenyon} Kenyon, S. L. \& Storey,
	J.W.V., 2005, \pasp, in press, astro-ph/0511510
\bibitem[1991]{ks} Krisciunas, K. \& Schaefer,
        B.E. 1991, \pasp, 103, 1033
\bibitem[1992]{landolt} Landolt, A. U. 1992, \aj, 104, 340
\bibitem[2004]{lawrence} Lawrence, J.S., Ashley, M.C.B., Tokovinin, A. \&
	Travouillon, T. 2004, Nature, 431, 278
\bibitem[2005]{mateshvili} Mateshvili, N., Fussen, D., Vanhellemont, F.,
	Bingen, C., Kyr\"l\"a, E., Mateshvili, I. \& Mateshvili, G., 2005,
        J. Geophys. Res., 110, D09209
\bibitem[1976]{mccartney} McCartney, E.J., 1976, Optics of the Atmosphere,
	(New York, John Wiley \& Sons)
\bibitem[1999]{oug} Ougolnikov, O.S. 1999, Cosmic Research,
        37, n.2, 159
\bibitem[2002]{oug02} Ougolnikov, O.S. \& Maslov, I.A., 2002, Cosmic Research,
	40, 224
\bibitem[2004]{oug04} Ugolnikov, O.S, Postylyakov, O.V. \& Maslov, I.A.,
	2004, J. Quant. Spec. Radiat. Transf., 88, 233
\bibitem[2003a]{patat03a} Patat, F., 2003a, A\&A, 400, 1183 
\bibitem[2003b]{patat03b} Patat, F., 2003b, A\&A, 401, 797 
\bibitem[2004]{postylyakov} Postylyakov, O.V., 2004, J. Quant. Spec. Radiat. 
	Transf., 88, 297
(the same issue with paper by Ugolnikov, Postylyakov, Maslov).
\bibitem[1966]{rozenberg} Rozenberg, G.V. 1966, Twilight,
        (New York, Plenum Press)
\bibitem[1977]{smart} Smart, W.M., 1977, Textbook on Spherical Astronomy,
	Sixth Edition, (Cambridge, Cambridge University Press)
\bibitem[2002]{szeifert} Szeifert, T., 2002, FORS1+2 User's Manual, 
	VLT-MAN-ESO-13100-1543, Issue 2.3
\bibitem[1993]{tyson} Tyson, N.D. \& Gal, R. 1993, \aj, 105, 1206
\bibitem[1988]{wu} Wu, B. \& Lu, D., 1988, Applied Optics, 27, 4899
\end{thebibliography}
\end{document}